\documentclass[aps,prl,twocolumn,groupedaddress]{revtex4}

\usepackage{graphicx}
\usepackage{amsmath}
\usepackage{relsize}
\usepackage{hyperref}

\begin{document}

\title{Coexistence of the superconducting energy gap and pseudogap {\it above} and below the transition temperature of superconducting cuprates}

\author{J. L. Tallon$^{1,2}$, F. Barber$^{1}$, J. G. Storey$^{2}$ and J. W. Loram$^{2}$}

\affiliation{$^1$MacDiarmid Institute, Industrial Research Ltd.,
P.O. Box 31310, Lower Hutt, New Zealand.}

\affiliation{$^2$Cavendish Laboratory, Cambridge University, CB3 0HE, United Kingdom.}

\begin{abstract}
We express the superconducting gap, $\Delta(T)$, in terms of thermodynamic functions in both $s$- and $d$-wave symmetries. Applying to Bi$_2$Sr$_2$CaCu$_2$O$_{8+\delta}$ and Y$_{0.8}$Ca$_{0.2}$Ba$_2$Cu$_3$O$_{7-\delta}$ we find that for all dopings $\Delta(T)$ persists, as a partial gap, high above $T_c$ due to strong superconducting fluctuations. Therefore in general two gaps are present above $T_c$, the superconducting gap and the pseudogap, effectively reconciling two highly polarized views concerning pseudogap physics.
\end{abstract}


\maketitle

On cooling a superconductor (SC) below $T_c$ coherent pairing of electrons 
opens a gap, $\Delta$, centered at the Fermi level. In a conventional SC $\Delta(T)$ closes at $T_c$ but for underdoped cuprates a partial gap is found to persist above $T_c$ and this is widely attributed to the so-called {\it pseudogap} \cite{Norman}. The field is sharply divided as to the origins of the pseudogap. One view is that it is some form of precursor SC state while another is that it arises from some correlation that competes with the SC state \cite{Norman}, so that the two gaps coexist below $T_c$. The inherent physics for each scenario is fundamentally different. In the former case a phase-incoherent SC state \cite{Emery} emerging from RVB physics high above $T_c$ \cite{Anderson} is often invoked, implying a very large SC energy gap which falls rapidly with increasing doping. In the latter case, it is the pseudogap, arising from some independent competing correlation, that has the large energy scale and the pseudogap closes abruptly at a putative ground-state quantum critical point lying within the SC dome at $p_{crit}$=0.19 holes/Cu \cite{TallonLoram}.

Because these two scenarios differ so radically it remains a central challenge to identify the nature of these energy gaps. Is there, indeed, one or two distinct gaps? Here we present a new method to calculate $\Delta(T)$ from the electronic specific heat. We show that, for the cuprates at any doping, $\Delta$ appears to remain finite above $T_c$ reflecting a partial gap arising from strong SC fluctuations and, in the underdoped region, coexists there with the pseudogap. Thus, in a sense, {\it both scenarios prove to be correct}. There are two gaps above $T_c$ just as there are two gaps below $T_c$ so that both fluctuations and competing pseudogap correlations play a key role in HTS physics.

Using a high-resolution differential technique Loram {\it et al.} \cite{Loram} have been able to isolate the electronic specific heat from the much larger phonon term in a number of high-$T_c$ cuprates. This has allowed many important conclusions to be drawn \cite{Loram}, including the fact that, due to strong SC fluctuations, the mean-field (MF) transition temperature, $T_c^{mf}$, determined from entropy conservation, lies well above the observed $T_c$ value (by up to 50K) \cite{Fluc,Wen}. Hereafter, we drop the descriptor `electronic' and by the terms specific heat, $C_P$, specific heat coefficient, $\gamma \equiv C_P/T$, entropy, $S$, internal energy, $U$, and free energy, $F$, we mean the electronic components of these.

We draw largely on Ferrell \cite{Ferrell} and extend to include $d$-wave SC. Starting from the BCS Hamiltonian he shows:
\begin{equation}
\left(\frac{\partial F}{\partial T_c}\right)_T = - \zeta \alpha^2 N(0)T_cQ(t) \label{dFdTc}
\end{equation}
\noindent where $N(0)$ is the DOS at the Fermi level, $t \equiv T/T_c$, $Q(t) \equiv \left(\Delta(T)/\Delta_0\right)^2$ and we include the additional factor $\zeta=1$. For an anisotropic gap we take $\Delta$ to be the amplitude of the $\textbf{k}$-dependent gap. In this case Ferrell's $\Delta(T)^2$ should be replaced by a Fermi surface average $\langle\Delta_\textbf{k}(T)^2\rangle=\zeta\Delta(T)^2$ where $\zeta=1$ for $s$-wave and 1/2 for $d$-wave.  The BCS gap ratio $\alpha \equiv \Delta_0/k_BT_c = (\pi/\gamma_E)e^C$ where $\gamma_E=1.781...$ is Euler's constant and $C = 0$ for $s$-wave while $C=\ln2-1/2$ for $d$-wave \cite{Maki}. Ferrell then integrates Eq.~\ref{dFdTc} over all $T_c > T$ to effectively obtain:
\begin{align}
\Delta F(T) &= F_n(T)-F_s(T,T_c) \nonumber \\
&= \zeta N(0)\Delta_0^2\,t^2 \int_t^1 \! t^{\prime-3} \, Q(t^{\prime}) \,\mathrm{d}t^{\prime}
\label{deltaF}
\end{align}

\begin{figure}
\centerline{\includegraphics*[width=75mm]{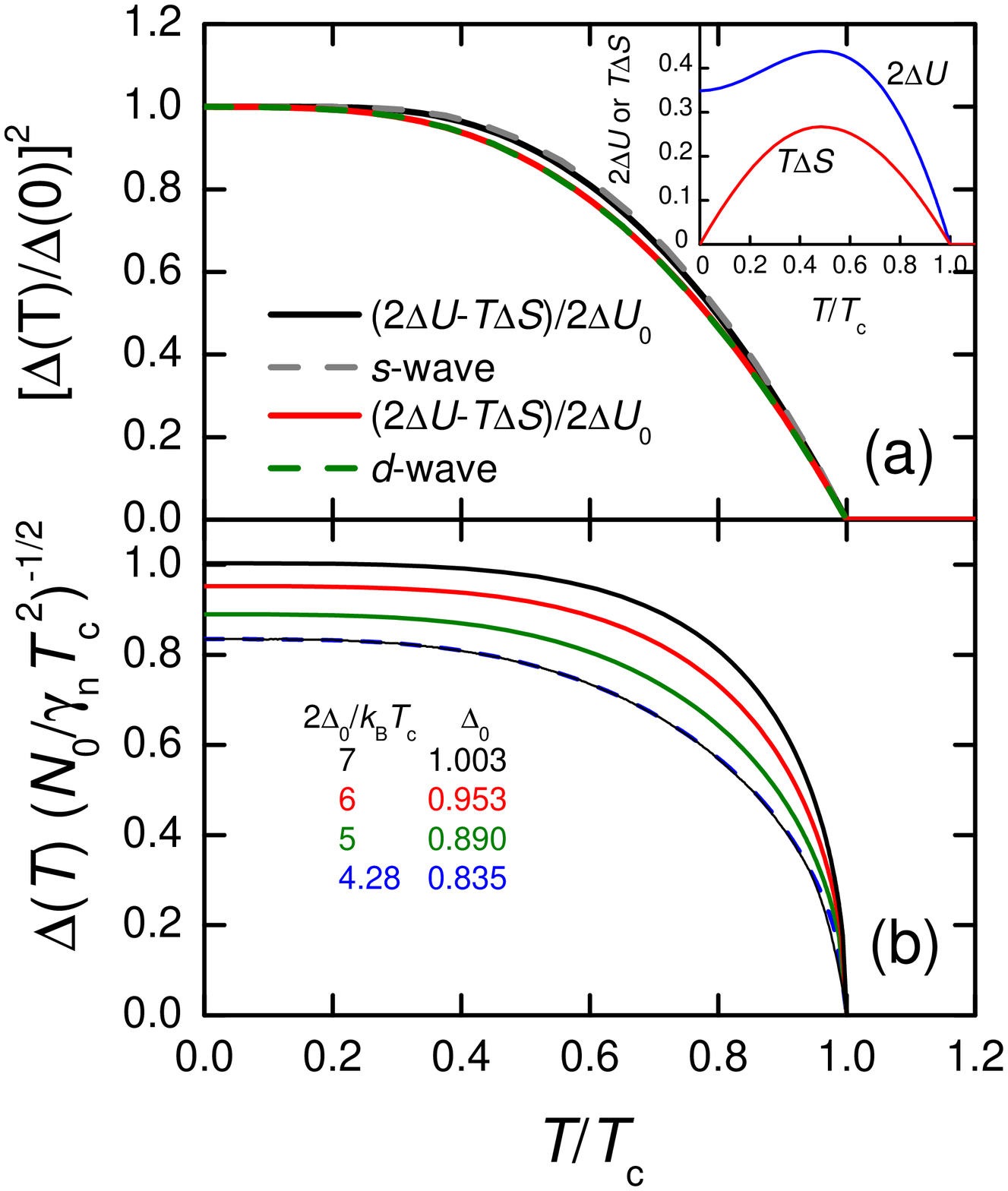}} \caption{\small
(Color online) (a) The temperature dependence of the square of the normalized SC gap function $(\Delta(T)/\Delta_0)^2$ for $s$-wave and $d$-wave weak-coupling BCS (dashed curves). These are compared with the values of this parameter calculated from Eq,~\ref{energygap} using $(2\Delta U(T) - T\Delta S(T))/2\Delta U(0)$. The inset shows the $T$-dependence of each contribution $2\Delta U$ and $T\Delta S$.
(b) The $T$-dependence of the SC gap function $\Delta(T)$ calculated using Eq.~\ref{energygap} for $d$-wave strong-coupling BCS with $2\alpha = 4.28$ (weak coupling), 5, 6 and 7.}
\label{BCSgaps}
\end{figure}

Ferrell's intention was to adopt a model $T$ dependence of $Q(t)$ from which to calculate $\Delta F(T)$. Our task is the opposite, to calculate $\Delta(T)$ from $\Delta F(T)$ derived from specific-heat data. By differentiating each side of Eq.~\ref{deltaF} with respect to $T$ and rearranging we obtain
\begin{align}
\zeta N(0) \Delta(T)^2 &= 2\Delta F(T) + T\Delta S(T) \nonumber \\
&\equiv 2\Delta U(T) - T\Delta S(T)
\label{energygap}
\end{align}
\noindent which expresses $\Delta(T)$ directly in terms of thermodynamic functions $\Delta F$, $\Delta S = S_n - S_s$ and $\Delta U = U_n - U_s$.

Quite generally, for a second-order MF phase transition near $T_c$, $\Delta F(T) = -\frac{1}{2} \Delta \gamma_c (T-T_c)^2$, so
\begin{equation}
\Delta(T)^2 \rightarrow \frac{2T_c^2 \Delta\gamma_c}{\zeta N(0)} \left(1 - T/T_c\right) \label{DatTc}
\end{equation}
\noindent where $\Delta\gamma_c$ is the jump in $\gamma$ at $T_c$. This means that the coherence length $\xi(T) = \hbar V_F/\pi \Delta(T)$ has the correct $(1-t)^{-1/2}$ dependence near $T_c$.

We have computed $(2\Delta U(T) - T\Delta S(T))/2\Delta U(0)$ for both $s$- and $d$-wave weak-coupling BCS and in Fig.~\ref{BCSgaps}(a) we compare these (solid curves) with the theoretical $T$-dependence of $(\Delta(T)/\Delta_0)^2$ (dashed curves). For both symmetries there is excellent agreement across the entire $T$-range and the gap amplitude satisfies:
\begin{equation}
\Delta F(0) = \Delta U(0) \equiv U_0 = \frac{1}{2} \zeta N(0) \Delta_0^2 . \label{U0}
\end{equation}
\noindent This is just the ground-state condensation energy. The inset shows the individual contributions $2\Delta U$ and $T\Delta S$ to $\Delta(T)$. $\Delta U(T)$ passes through a maximum while subtraction of the entropy term recovers the canonical monotonic $T$-dependence of the $s$- or $d$-wave gap.

Ferrell's  theory is strictly for weak-coupling BCS, based on the logarithmic relation between the pairing interaction and $T_c$. Extending to strong-coupling we may employ the Padamsee $\alpha$-model approximation \cite{Padamsee} where the ratio $\alpha \equiv \Delta_0/k_BT_c$ is the only adjustable parameter and $\Delta(T)/\Delta_0$ is assumed to follow the weak-coupling BCS form for all $\alpha$. As $\alpha$ cancels in Eq.~\ref{deltaF} we might still consider using Eq.~\ref{energygap} to calculate $\Delta(T)$. In the case of Pb, a strong-coupling superconductor, we have calculated $\Delta F$ and $\Delta S$ from critical-field measurements, and thence $\Delta(T)$ using Eq.~\ref{energygap}. We find excellent agreement with measurements of the gap from tunneling including the flattening of $\Delta(T)$ relative to the BCS $T$-dependence. This gives us confidence to extend beyond weak-coupling, as may be necessary for the cuprates.

Accordingly, we used the $\alpha$-model to calculate $\Delta F$ and $\Delta S$ for $2\alpha = 4.28$ (weak coupling), 5, 6 and 7 in a $d$-wave scenario, employing the same method as Padamsee {\it et al.} \cite{Padamsee}. Fig.~\ref{BCSgaps}(b) shows $\Delta(T)$ calculated for each case using Eq.~\ref{energygap}. The fine black curve under the blue dashed curve for $2\alpha = 4.28$ is the theoretical weak-coupling BCS gap, for which the match is exact. In the figure $\gamma_n$ is the normal-state (NS) value of $\gamma$, which is assumed to be $T$-independent. In strong coupling, $\gamma_n$ is enhanced by a factor $(1+\lambda)$ above its Sommerfeld value, viz.
\begin{equation}
\gamma_n = \frac{2}{3}\pi^2k_B^2N_0(1+\lambda) , \label{enhancement}
\end{equation}
\noindent where $\lambda$ is the usual electron-boson coupling parameter in Eliashberg theory \cite{Carbotte} and $N_0$ is the bare band DOS, un-renormalized by electron-boson or Coulomb effects. Thus $\Delta(T)$ in Fig.~\ref{BCSgaps}(b) is expressed in units of $\sqrt{\frac{2}{3}} \pi k_BT_c \sqrt{1+\lambda}$. 
Leaving aside the absolute magnitude of $\Delta$, the $T$-dependence of $\Delta$ evidently flattens with increasing coupling. Though this violates the main premise of the $\alpha$-model, the $\alpha$-model could potentially be refined by calculating $\Delta(T)$ iteratively. For the time being, Eq.~\ref{energygap} seems to be a satisfactory approximation for strong coupling if $2\alpha$$<$5, as we find for the cuprates \cite{SOM}.

\begin{figure}
\centerline{\includegraphics*[width=75mm]{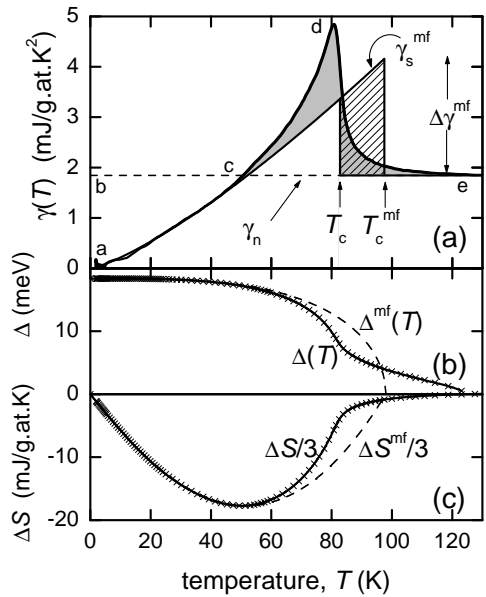}} \caption{\small
(Color online) (a) Reproduced from \cite{Fluc}: analysis of the specific heat coefficient, $\gamma(T)$, for Y$_{0.8}$Ca$_{0.2}$Ba$_2$Cu$_3$O$_{6.75}$ to determine the mean-field $T_c$ value, $T_c^{mf}$, showing the deduced MF coefficient $\gamma^{mf}$ and the symmetric fluctuation contribution (grey shading). By entropy balance the hatched area equals the shaded area under the fluctuation term. (b) solid curve: the SC energy gap $\Delta_0$ calculated using Eq.~\ref{energygap} and dashed curve: its MF value, $\Delta_0^{mf}$ calculated from $\gamma^{mf}$ in (a). (c) solid curve: the entropy difference $\Delta S = S_s - S_n$ calculated by integrating $(\gamma - \gamma_n)$ from (a); dashed curve $\Delta S^{mf}$ calculated by integrating $(\gamma^{mf} - \gamma_n)$ from (a). }
\label{MFgap}
\end{figure}

We now apply this analysis to the electronic specific heat of the HTS cuprates reported by Loram {\it et al.} \cite{Loram}. Fig.~\ref{MFgap}(a) shows the previously reported analysis \cite{Fluc} of the specific heat coefficient, $\gamma(T)$, for Y$_{0.8}$Ca$_{0.2}$Ba$_2$Cu$_3$O$_{6.75}$ used to determine the mean-field $T_c$ value, $T_c^{mf}$. At this doping ($p=0.185$) the pseudogap is absent and the NS coefficient $\gamma_n(T)$ is essentially constant (dashed line). $\gamma_s^{mf}$ is the MF $\gamma$ in the SC state deduced by entropy balance, namely the area $abc$ equals the area $cde$. Also by entropy balance the grey shaded area under the fluctuation contribution equals the hatched area which therefore defines $T_c^{mf}$. By integrating $\gamma(T) - \gamma_n(T)$ in Fig.~\ref{MFgap}(a) we obtain $\Delta S = S_s - S_n$ and similarly $\Delta S^{mf} = \int_0^T (\gamma_s^{mf} - \gamma_n)dT $. These are plotted in Fig.~\ref{MFgap}(c) by the solid and dashed curves, respectively, where only every 4$^\text{th}$ data point is shown. These may in turn be integrated to generate $\Delta F(T)$ and $\Delta F^{mf}(T)$ and these combined with $T\Delta S(T)$ and $T\Delta S^{mf}(T)$ to generate $\Delta(T)$ and $\Delta^{mf}(T)$ using Eq.~\ref{energygap}. These are plotted in Fig.~\ref{MFgap}(b) where $N(0)$ is obtained using Eq.~\ref{enhancement} with $\lambda =0$. The actual gap should be larger by the factor $\sqrt{1+\lambda}$.

The first point to note is that $\Delta^{mf}(T)$ is found to follow almost precisely the BCS temperature dependence. This means that the HTS systems are close to weak-coupling behavior as we have previously deduced \cite{Fluc} thus justifying the basic assumptions of our analysis. Even if $\lambda$ is appreciable one could invoke the Padamsee approach to renormalize the magnitude of $\Delta$ provided $2\Delta/k_BT_c^{mf}$ does not greatly exceed the BCS value of 4.3. Secondly, with increasing temperature $\Delta(T)$ starts to fall below $\Delta^{mf}(T)$ at the onset of SC fluctuations below $T_c$. At $T_c$ there is an inflexion in $\Delta(T)$ which then remains finite and falls only slowly to zero above $T_c$. As it does so it becomes less well defined due to the square root in Eq.~\ref{energygap}. At $T_c$ the coherent SC state vanishes and this finite residual ``gap" reflects a fluctuation-induced loss, above $T_c$, of spectral weight in the DOS at $E_F$ - just as described by  Fig. 10.2 in Larkin and Varlamov \cite{Larkin}.

\begin{figure}
\centerline{\includegraphics*[width=75mm]{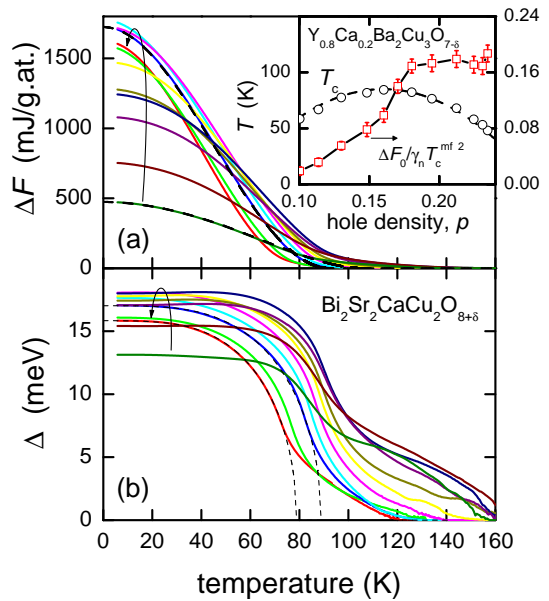}} \caption{\small
(Color online) The $T$-dependence of (a) the condensation free energy $\Delta F(T)$ and (b) $\Delta$ obtained using Eq.~\ref{energygap}, for Bi$_2$Sr$_2$CaCu$_2$O$_{8+\delta}$. Curved arrows show increasing doping from $p=0.12$ to $p=0.22$. Note that the absolute value of $\Delta(T)$ is larger than that shown by a factor of $\sqrt{1+\lambda}$. Inset: the doping dependence of the BCS ratio $\Delta F(0)/\gamma_n (T_c^{mf})^2$ determined for Y$_{0.8}$Ca$_{0.2}$Ba$_2$Cu$_3$O$_{7-\delta}$. The BCS value of 0.17 is preserved for $p\geq0.19$ but falls rapidly with the opening of the pseudogap. }
\label{calcgaps}
\end{figure}

A similar analysis was carried out for many other doping levels for both Y$_{0.8}$Ca$_{0.2}$Ba$_2$Cu$_3$O$_{7-\delta}$ and Bi$_2$Sr$_2$CaCu$_2$O$_{8+\delta}$. Now we must take into account the effects of the proximate van Hove singularity (vHs) on the overdoped side ($\gamma_n(T)$ rises with decreasing $T$) and the pseudogap on the underdoped side ($\gamma_n(T)$ falls with decreasing $T$). To do this we still use entropy balance but employ a rigid ARPES-derived dispersion, which implicitly contains the vHs, to determine the doping evolution of the background $\gamma_n(T)$ and $S_n(T)$. We use the model of Storey {\it et al} \cite{StoreyFArc} which includes a non-nodal NS pseudogap at ($\pi$,0) reflecting the formation of hole pockets as described, for example, by the Fermi-surface-reconstruction model of Yang, Rice and Zhang \cite{Yang}. The pseudogap closes abruptly at $p\approx0.19$. For more details see SM \cite{SOM}. For Bi$_2$Sr$_2$CaCu$_2$O$_{8+\delta}$ the data for $S(T)$ and the dispersion-derived $S_n(T)$ are already reported by Storey {\it et al.} \cite{StoreyFArc}. By integrating $\Delta S(T) = S(T)-S_n(T)$ we obtain $\Delta F(T)$ which is shown in Fig.~\ref{calcgaps}(a) for 11 dopings spanning under- and over-doped regions.

\begin{figure*}
\centering
\includegraphics[width=130mm]{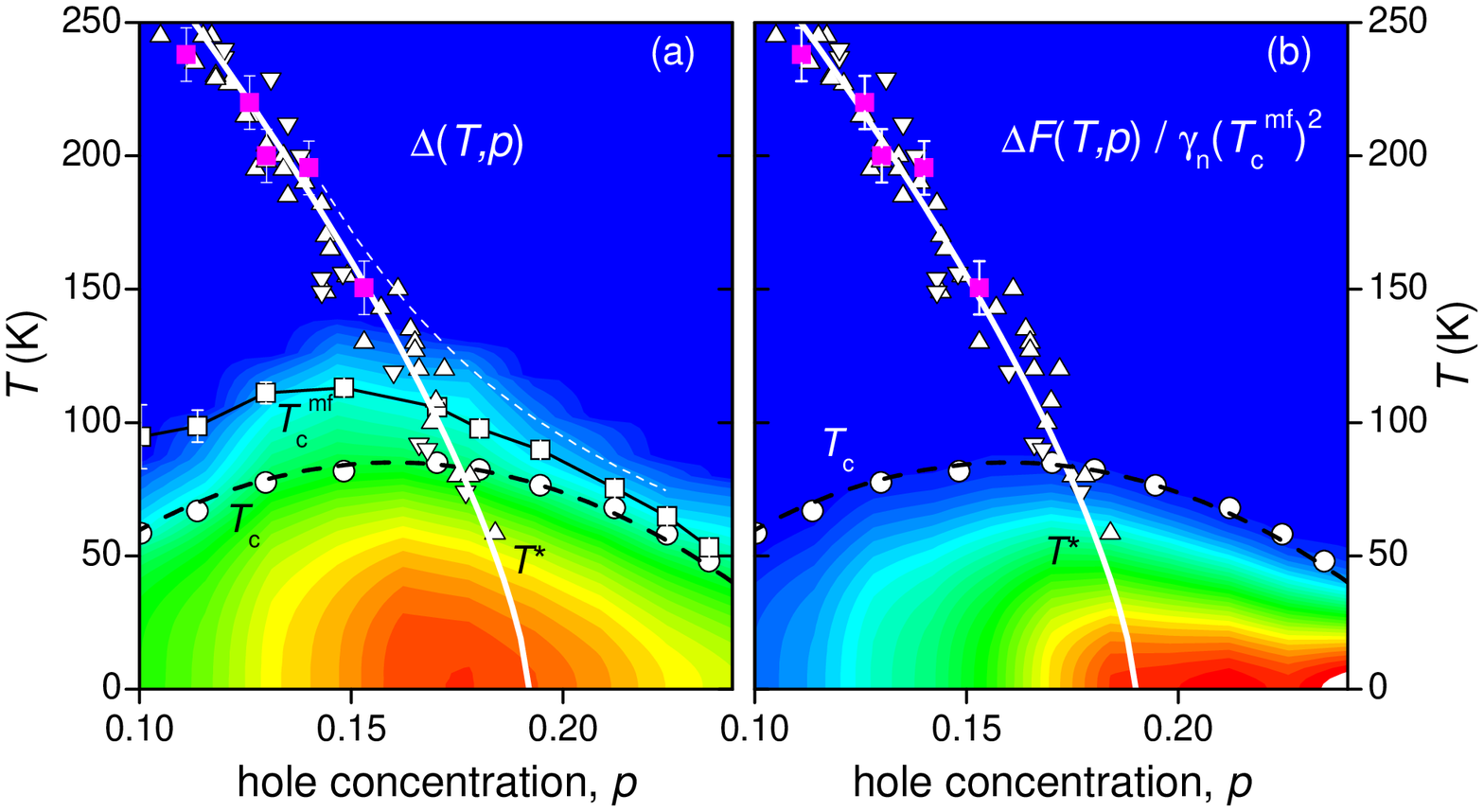}
\caption{\small
(Color online) False color plot of (a) the SC gap, $\Delta(T,p)$ and (b) the BCS-normalised condensation free energy $\Delta F(T)/\gamma_n (T_c^{mf})^2$ across the $(p-T)$ phase diagram for Y$_{0.8}$Ca$_{0.2}$Ba$_2$Cu$_3$O$_{7-\delta}$. In (a) the color scale runs from 20 meV (red) to 0 meV (blue) and in (b) from 0.18 (red) to 0 (blue). Also shown is the observed $T_c$ and $T_c^{mf}$ determined previously \cite{Fluc}. The SC gap extends well above $T_c$ while $\Delta F(T)$ is cut off at $T_c$. Also shown is $T^*(p)$ (white line) previously reported for Y$_{1-x}$Ca$_{x}$Ba$_2$Cu$_3$O$_{7-\delta}$ (epitaxial thin films: down-triangles; polycrystalline: up-triangles) and also as reported by Daou {\it et al.} \cite{Daou} (magenta squares) for YBa$_2$Cu$_3$O$_{7-\delta}$ single crystals. In (a) the dashed white curve is the envelope of the pseudogap in the underdoped region and the residual SC gap in the overdoped region. In (b) $\Delta F(0)/\gamma_n (T_c^{mf})^2$ adopts the BCS value 0.17 across the overdoped region but collapses rapidly at $T^*$ confirming that the $T^*$ line does indeed terminate at $p=0.19$. }
\label{Delta123}
\end{figure*}

From $\Delta F(T,p)$ and $T\Delta S(T,p)$ we calculate $\Delta(T,p)$ using Eq.~\ref{energygap}. This is plotted in Fig.~\ref{calcgaps}(b) and as before the deduced gap does not vanish at $T_c$. Rather, it inflects there and then persists some 20 K above $T_c$ in overdoped samples and up to 70 K above $T_c$ for underdoped samples. Residual gaps above $T_c$ are perhaps not new however they are usually confused with the pseudogap \cite{Renner1}. We distinguish between the two gaps, as follows.

The residual gap that we observe above $T_c$ arises from SC fluctuations near $T_c$ which are distinguished by a fluctuation term in $\gamma(T)$ which is symmetric over a narrow range about $T_c$ (see grey shaded areas in Fig.~\ref{MFgap}(a)). The pseudogap is altogether different. Its effects are not centered on $T_c$ but extend over a broad temperature range up to 300 K or more, and is distinguished by:

(i) a broad suppression of $S(T)/T$ as $T$ is reduced \cite{Loram,SOM}, corresponding precisely to the suppression of the spin susceptibility, $\chi_s(T)$, long observed in NMR \cite{Alloul89}.

(ii) the abrupt reduction in the jump $\Delta\gamma_c$ at $T_c$ with the opening of the pseudogap at $p=0.19$ holes/Cu. As $p$ is reduced below 0.19 $\Delta\gamma_c$ is rapidly diminished, reflecting a crossover from strong to weak superconductivity.

(iii) a relative insensitivity to the effect of a magnetic field or impurities \cite{Naqib,Alloul} in distinct contrast to the pairing gap arising from SC fluctuations.

As $\Delta(T)$ persists above $T_c$, even in overdoped samples where the pseudogap is absent, it must therefore arise from SC fluctuations above $T_c$. On theoretical \cite{Larkin} and experimental \cite{Dubroka} grounds, this will cause a gap-like loss of spectral weight and an associated entropy loss which underlies the residual $\Delta(T)$. In further support, Gomes {\it et al}. \cite{Gomes} observe a spatially inhomogeneous partial gap above $T_c$ in tunneling spectroscopy up to a temperature $T_{p,max}$ which closely matches our $T_c^{mf}$. Such a gap is also seen in ARPES \cite{Kondo}. This partial gap also probably underlies the anomalous Nernst effect observed in under- and over-doped samples between $T_c$ and $T_c^{mf}$ \cite{Ong1}.

Returning to Y$_{0.8}$Ca$_{0.2}$Ba$_2$Cu$_3$O$_{7-\delta}$, similar results are found. Fig.~\ref{Delta123}(a) shows a false-color plot of the magnitude of $\Delta(T,p)$ across the $p-T$ phase diagram, along with $T_c(p)$ and the previously determined $T_c^{mf}(p)$ \cite{Fluc}. A finite gap extends above $T_c$ and indeed above $T_c^{mf}$, reflecting again the presence of strong SC fluctuations, though neither the gap nor $T_c^{mf}$ extend as high as in the case of Bi$_2$Sr$_2$CaCu$_2$O$_{8+\delta}$. To emphasize the crucially important distinction between pseudogap and SC correlations we also plot in Fig.~\ref{Delta123}(a) previously determined $T^*$ values for Y$_{1-x}$Ca$_{x}$Ba$_2$Cu$_3$O$_{7-\delta}$ (epitaxial films: down-triangles, polycrystalline: up-triangles). $T^*$ was determined in the usual way by the downturn from linear resistivity, with the added precaution of using a magnetic field to distinguish between SC fluctuations and the pseudogap near $T_c$ \cite{Naqib,Alloul}. $T^*$ cuts through the crescent of finite $\Delta(T,p)$ above $T_c$, falling to zero at $p\approx0.19$.

This issue continues to be debated. Many groups espouse a $T^*$ line similar to the white dashed curve in Fig.~\ref{Delta123}(a) that extends above the SC dome, across the overdoped region. Daou {\it et al.} \cite{Daou} are a recent example. We therefore also plot Daou's $T^*$ data points (magenta squares) for YBa$_2$Cu$_3$O$_{7-\delta}$ and they are in excellent agreement with our data shown by the white triangles and solid white curve. The white dashed curve is the envelope above $T_c$ of a finite gap-like feature, whether the SC gap or the pseudogap. Unless these gaps are distinguished it is not surprising that many groups have failed to see that $T^*$ terminates abruptly at $p\approx0.19$.

Definitive evidence for the termination of $T^*(p)$ at $p=0.19$ is shown in Fig.~\ref{Delta123}(b). Here we plot a false color plot of the ratio $\Delta F(T)/\gamma_n (T_c^{mf})^2$ across the phase diagram. Note that we have used $T_c^{mf}$ and not $T_c$ as the normalising energy scale. The value of this ratio at $T=0$ is also plotted in the inset to Fig.~\ref{calcgaps}(a). The universal BCS $d$-wave value for this ratio is 0.17 and, indeed, this value is obtained across the entire overdoped region for $p\geq0.19$. But with the opening of the pseudogap for $p<0.19$ the ratio is seen in Fig.~\ref{Delta123}(b) to collapse abruptly, clearly delineating the termination of the pseudogap $T^*$ line. Moreover, this shows that the pseudogap and superconducting gap coexist below $T_c$ as they do above $T_c$. We are thus obliged to conclude that $T^*(p)$ cuts the SC dome and terminates at $p\approx0.19$, contrary to the inference of Daou {\it et al.} \cite{Daou} though in fact their data is fully consistent with our scenario.

Finally, the magnitude of $\Delta_0$ obtained here is a little lower than that observed previously by e.g. infrared measurements\cite{Yu} where the amplitude is about 25 meV. But recall that $\Delta(T)$ in Figs. ~\ref{calcgaps} and ~\ref{Delta123} is yet to be enhanced by the factor $\sqrt{1+\lambda}$.

In summary, we have shown that the SC gap, $\Delta(T)$, may be calculated from the electronic specific heat and we apply to the cuprates. For all dopings a residual finite $\Delta(T)$ extends up to 70K above $T_c$ reflecting a fluctuation-induced loss of spectral weight at $E_F$. This crescent of residual SC gap above $T_c$ is cut by the $T^*$ line showing that two gap-like features are present above $T_c$, one extending across the entire SC phase diagram due to strong SC fluctuations and the other present only in the optimal and underdoped region due to  the pseudogap. The ratio $U_0/\gamma_n (T_c^{mf})^2$ adopts the BCS weak coupling value (0.17) across the entire overdoped region down to $p\approx0.19$ where the pseudogap opens and the ratio then collapses rapidly, thus exposing an abrupt crossover to ``weak" superconductivity as the Fermi surface reconstructs.

\end{document}